\pdfoutput=1
\documentclass{PoS}
\usepackage{amssymb,amsfonts,amsmath}

\title{QCD at non-zero temperature and magnetic field}

\ShortTitle{QCD at non-zero temperature ...}

\author{\speaker{Kalman Szabo}\\
        Bergische Universit\"at, D-42119 Wuppertal, Germany\\
	J\"ulich Supercomputing Centre, Forschungszentrum J\"ulich, D-52425 J\"ulich, Germany\\
        E-mail: \email{szaboka@general.elte.hu}}

\abstract{A status of lattice QCD thermodynamics, as of 2013, is summarized.
Only bulk thermodynamics is considered. There is a separate section on magnetic
fields.}

\FullConference{31st International Symposium on Lattice Field Theory - LATTICE 2013\\
		July 29 - August 3, 2013\\
		Mainz, Germany}

\begin{document}

\section{Crossover, transition temperature and equation of state}

Three results in lattice QCD thermodynamics matured to a well-established
status in recent years: the nature of the transition, the transition
temperature and the equation of state. This means, that all systematic
uncertainties have either been eliminated or are kept under control. The
computations use physical values for the quark masses, a continuum
extrapolation is carried out from upto five different lattice spacings and
different physical volumes are used to control the finite size error.  The only
point of concern, that all these results are obtained with staggered fermions.
Though there is currently no sign, that in these observables the use of
staggered fermions would cause any problems, it is highly desirable to have
cross-checks using other fermion formulations.

\subsection{Crossover}

\begin{figure}
\begin{center}
\includegraphics[width=0.45\textwidth,viewport=400 10 563 154,clip]{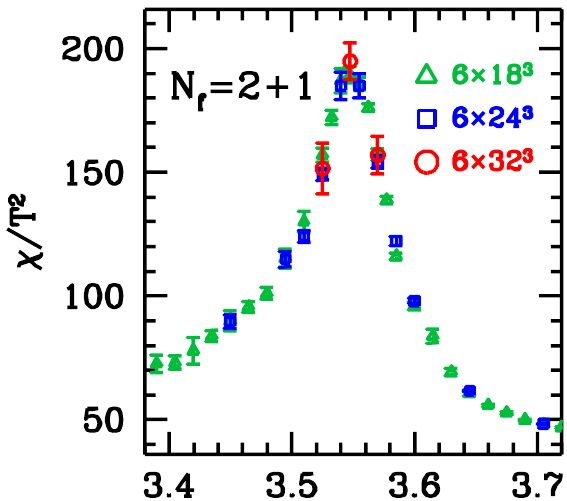}
\caption{The chiral susceptibility does not depend on the volume. It means, that
the QCD transition is a crossover \cite{Aoki:2006we}.}
\label{fi:co}
\end{center}
\end{figure}

The transition from the low temperature hadron dominated phase to the high
temperature quark-gluon plasma is a crossover \cite{Aoki:2006we}: there is no singularity in the
transition region. The chiral susceptibility peak is shown on Figure \ref{fi:co} in the
transition region for three different volumes. The height and the width of the
peak remains constant, as one increases the volume, no diverging behaviour is
seen, which is the characteristic of a crossover transition.

\subsection{Transition temperature}
\label{se:tc}

In a crossover there is no single critical temperature, which would separate
the two phases of matter. Transition temperatures, $T_c$'s can still be defined
as characteristic points (peak position or inflection point) of
observables.  Different observables can in general give different transition
temperature values, this is a feature of the crossover transition.  A
particularly interesting $T_c$ is the one defined as the peak position of the
chiral susceptibility. Its value was highly disputed, a consensus about its value
has only been reached recently:
\begin{align*}
T_c&= 147(2)(3) \text{ MeV}\quad \text{Wuppertal-Budapest (WB) group \cite{Aoki:2006br,Aoki:2009sc,Borsanyi:2010bp}},\\
T_c&= 154(9)    \text{ MeV}\quad\quad \text{ hotQCD collaboration \cite{Bazavov:2011nk}}.
\end{align*}

\begin{figure}
\begin{center}
\includegraphics[width=0.6\textwidth]{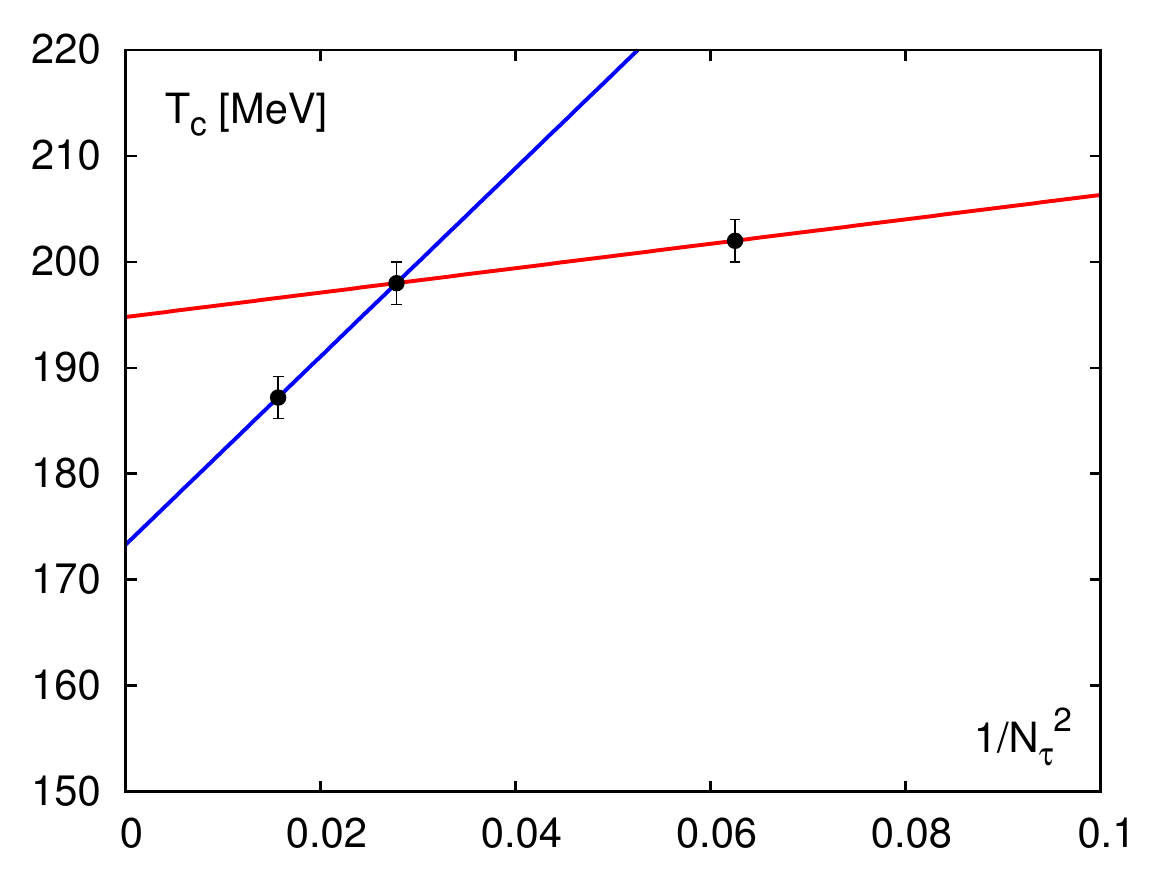}
\caption{The QCD transition temperature in early determinations. There is a strong decrease with the lattice spacing.
The now accepted continuum value is $T_c\sim150$ MeV. Figure from \cite{Bazavov:2011nk}.}
\label{fi:tc}
\end{center}
\end{figure}

Back in 2006 the Bielefeld-Brookhaven-Columbia-Riken collaboration, which later
merged with part of the MILC collaboration and formed the hotQCD, was reporting
considerably larger values for this transition temperature \cite{Cheng:2006qk}.
The reason is now widely accepted: the lattice artefacts on the lattices, that
were used in \cite{Cheng:2006qk}, were very large. This is demonstrated on
Figure \ref{fi:tc}. The 2006 continuum
extrapolation is the red line, which was carried out using the two coarsest
lattices and resulted in $T_c\sim 190$ MeV. Adding a finer lattice changed the
continuum extrapolation, this is the blue line, which decreased the value of
the transition temperature significantly.

Other fermion formulations are also used to determine $T_c$. Though even the
most advanced ones are not in the continuum, see the domain wall fermion result
of \cite{Buchoff:2013nra} or using larger than physical quark masses, see the
Wilson fermion result of \cite{Borsanyi:2012uq}.

\subsection{Equation of state}

\begin{figure}
\begin{center}
\includegraphics[width=0.60\textwidth]{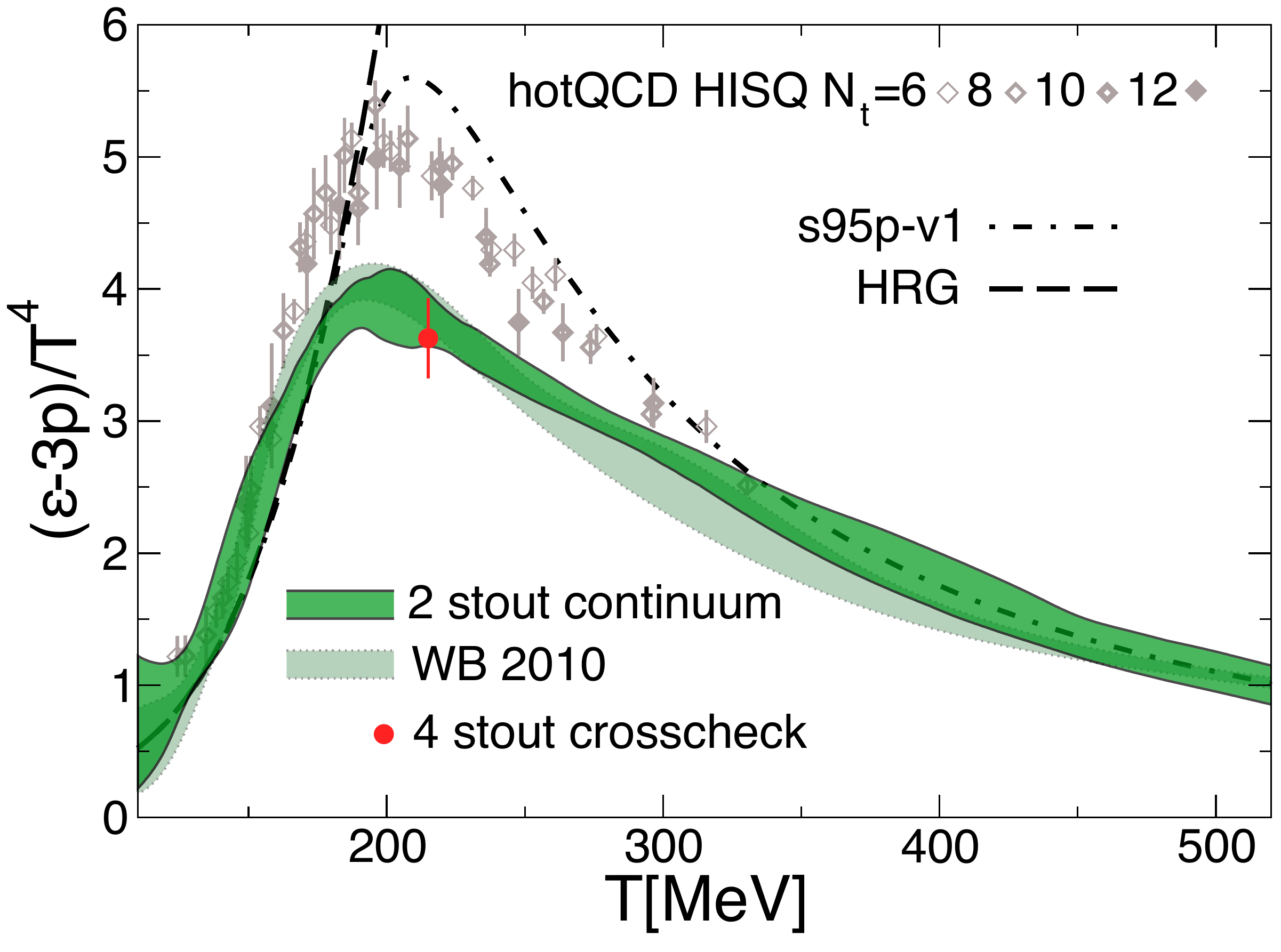}\\
\includegraphics[width=0.60\textwidth]{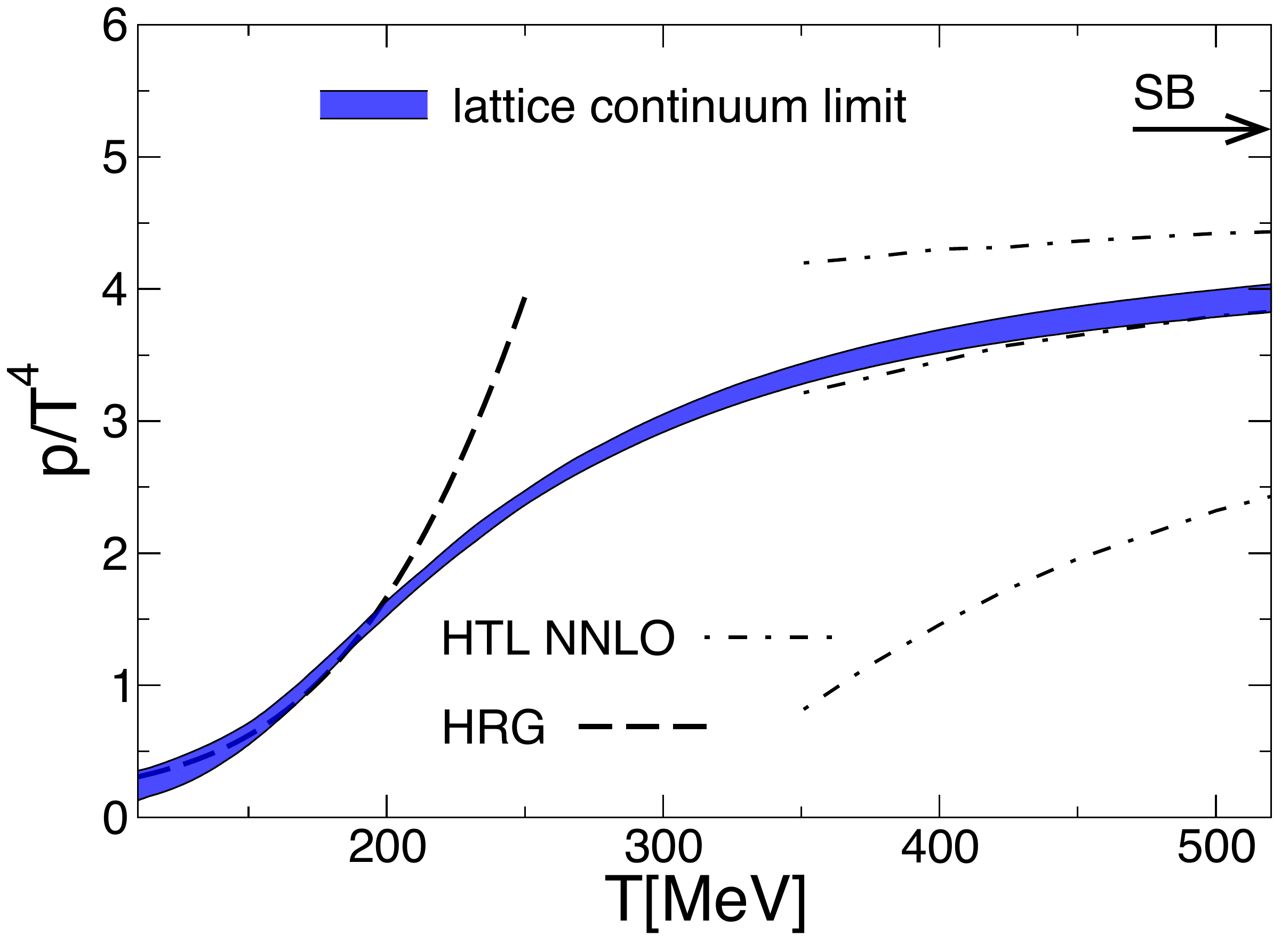}
\caption{The 2+1 flavor continuum extrapolated trace anomaly (up) and pressure (down).}
\label{fi:eos}
\end{center}
\end{figure}

In 2013 the 2+1 flavor equation of state has also become member of the ``the
continuum extrapolated results with physical quark masses'' club \cite{Borsanyi:2013bia}. The trace
anomaly and the pressure are shown on the upper and lower panel of Figure \ref{fi:eos}. Both
are consistent with the Hadron Resonance Gas model for small temperatures. The
following parameterization describes the trace anomaly as the function of the
temperature
\begin{align*}
\frac{I(T)}{T^4} = \exp(-h_1/t-h_2/t^2)\cdot\left( h_0 + \frac{f_0 [\tanh(f_1\cdot t+f_2)+1]}{1+g_1\cdot t + g_2 \cdot 
t^2} \right),
\end{align*}
where $t=T/200\text{ MeV}$ and the coefficients are:
\begin{center}
\begin{tabular}{|c|c|c|c|c|c|c|c|c|}
\hline
&$h_0$&$h_1$&$h_2$&$f_0$&$f_1$&$f_2$&$g_1$&$g_2$\\
\hline
2+1&0.1396 & -0.1800 & 0.0350 & 1.05 & 6.39 & -4.72 & -0.92 &0.57 \\
\hline
\end{tabular}
\end{center}

\noindent A full result in the 2+1 flavor case, although it neglects the charm
quark contribution was necessitated by a discrepancy: there was a $\sim$20\%
difference in the peak height of the trace anomaly between the hotQCD data and
the WB data from 2010 \cite{Borsanyi:2010cj}. The WB group has confirmed their
previous equation of state calculation by improving it in many ways, as was
described by Krieg in his talk \cite{Borsanyi:2013bia}. These include a
continuum extrapolation from five different lattice spacings, an improved
determination of the zero point of the pressure, an improved systematic error
determination and a cross-check with a somewhat different staggered action. The
updated WB results are in complete agreement with those from 2010. There was no
update from the hotQCD group, so the discrepancy still remains to be resolved.

A 2+1 flavor result cannot be considered as the final one, due to the omission
of the charm quark, which becomes important for high enough temperatures.  An
estimate of the 2+1+1 flavor trace anomaly, which is based on partial quenching
the charm quark, is given by the same formula as before with the following
parameters \cite{Borsanyi:2010cj}:
\begin{center}
\begin{tabular}{|c|c|c|c|c|c|c|c|c|}
\hline
&$h_0$&$h_1$&$h_2$&$f_0$&$f_1$&$f_2$&$g_1$&$g_2$\\
\hline
2+1+1&0.1396 & -0.1800 & 0.0350 & 5.59 & 7.34 & -5.60 & 1.42 & 0.50\\
\hline
\end{tabular}
\end{center}

\noindent Several groups are now carrying out simulations with a dynamical
charm quark. One of them is the MILC collaboration, who use highly improved
staggered quarks down to a pion mass of $m_\pi\sim$ 300 MeV, the results were
presented by Bazavov \cite{Bazavov:2013pra}. The tmfT collaboration utilizes
twisted mass quarks, down to a pion mass of $m_\pi\sim$ 400 MeV, as was described
by Burger \cite{Burger:2013hia}. Note, that although these simulations are done
at more than one lattice spacings, the results are not yet continuum
extrapolated.

An achievement of recent years, which is important from theoretical point of
view, is that lattice simulations can be run at such high temperatures, where a
connection to perturbation theory is possible. In pure Yang-Mills theory
this is now done \cite{Borsanyi:2012ve}.

\section{Fluctuations}

Baryon number ($B$), electric charge ($Q$) and strangeness ($S$) fluctuations have become
major topics in QCD thermodynamics. They are defined by differentiating the partition function
with respect to the baryon, charge and strange chemical potentials:
\begin{align*}
\chi^{BQS}_{ijk}=
\frac{1}{VT^3}
\left[\frac{\partial^i}{\partial (\mu_B/T)^i} \frac{\partial^j}{\partial (\mu_Q/T)^j} \frac{\partial^k}{\partial (\mu_S/T)^k}\right] \log Z.
\end{align*}
In the lattice community these quantities are better known as quark number
susceptibilities, they have been being calculated on the lattice since more
than a decade now.  The calculation poses significant numerical challenges: not
only the number of terms increases with the number of derivatives, but so does
the cancellation between these terms. The rule of thumb, that the volume
increases the statistics and helps reducing the noise does not apply to
fluctuations. For higher than second order fluctuations increasing the volume
actually makes the signal worse for a fixed number of configurations.
Currently there are continuum extrapolated results for all second order
fluctuations \cite{Borsanyi:2011sw,Bazavov:2012jq} and for some fourth order
ones \cite{Borsanyi:2013hza,Bellwied:2013cta}, including the baryon number
kurtosis. There also exist calculations for some of the sixth
\cite{Cheng:2008zh} and eight order cumulants \cite{Gavai:2008zr}, too. 

There are three main uses of fluctuations: exploring finite $\mu$ with Taylor
expansion, determining the dominant degrees of freedom of the system and
determining freezout parameters of heavy-ion collisions. Let us now discuss
them in detail.

\subsection{Finite $\mu_B$ with Taylor expansion}

Expanding an observable in a chemical potential results in expansion
coefficients, that can be computed with marginal effort, if the fluctuations
are already known. Thus knowing eg. baryon number fluctuations allows obtaining
results at $\mu_B>0$, which circumvents the infamous sign problem. Due to the
aforementioned numerical difficulties, this is only a solution for small
$\mu_B$'s. This technique made possible to determine important observables
related to finite $\mu_B$. The curvature of the transition line starting at
$\mu_B=0$ was determined in leading order in $\mu_B$:
\begin{align*}
\kappa= -T_c \left. \frac{ \partial T_c }{\partial \mu_B^2}\right|_{\mu_B=\mu_Q=0,\mu_S=-\mu_B/3}=
\begin{cases}
0.059(2)(4) & \text{BNL-Bielefeld based on $N_t=4,8$ \cite{Kaczmarek:2011zz},}\\
0.059(18)   & \text{WB continuum extrapolated \cite{Endrodi:2011gv}.}
\end{cases}
\end{align*}
Also the leading order $\mu_B$ correction to the equation of state at is known \cite{Borsanyi:2012cr}.

\subsection{Dominant degrees of freedom}

\begin{figure}
\begin{center}
\includegraphics[width=0.75\textwidth]{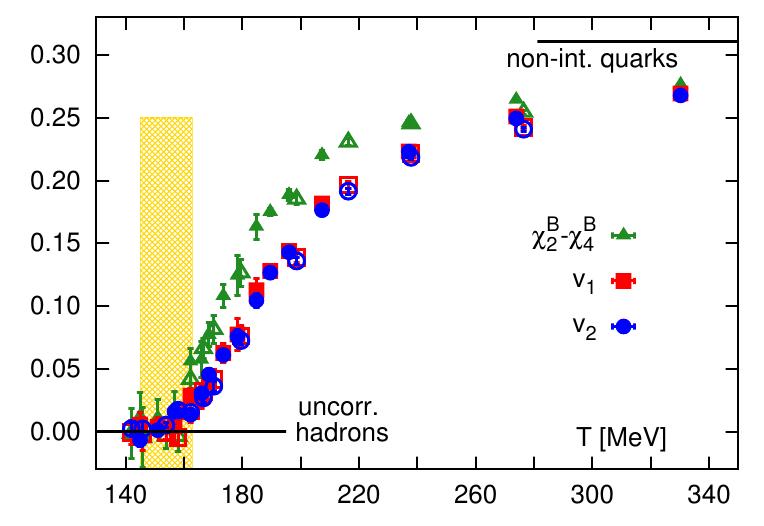}
\caption{The $v_1,v_2$ observables signalize the breakdown of the hadronic description \cite{Bazavov:2013dta}.}
\label{fi:v1}
\end{center}
\end{figure}
The sole value of a fluctuation is often used as an indicator to tell about the
dominant degrees of freedom in the system. For example let us consider the
following two combinations of fluctuations \cite{Bazavov:2013dta}:
\begin{align*}
v_1&=\chi_{11}^{BS} - \chi_{31}^{BS},\\
v_2&=(\chi_2^S - \chi_4^S)/3 + 2\chi_{13}^{BS} -4 \chi_{22}^{BS} + 2 \chi_{31}^{BS}.
\end{align*}
From the assumption, that the system is composed of uncorrelated particles, that carry
integer baryon number and strangeness, it follows, that $v_1$ and $v_2$ vanish.
One such model, that satisfies this condition is the standard Hadron Resonance
Gas. This observable can be determined in lattice QCD, see Figure \ref{fi:v1}. For small
temperatures it is zero supporting the assumption. For larger temperatures it
starts deviating from zero, so we conclude, that the system cannot be a
composition of uncorrelated particles carrying integer charges for
$T\gtrsim160$ MeV. For high temperatures similar observables can be
constructed. They can be used to test whether the dominant degrees of freedom
carrying strangeness have the same quantum numbers as free strange quarks.
These investigations were reported by Schmidt \cite{Schmidt:2013nra}.

\begin{figure}
\begin{center}
\includegraphics[width=0.60\textwidth]{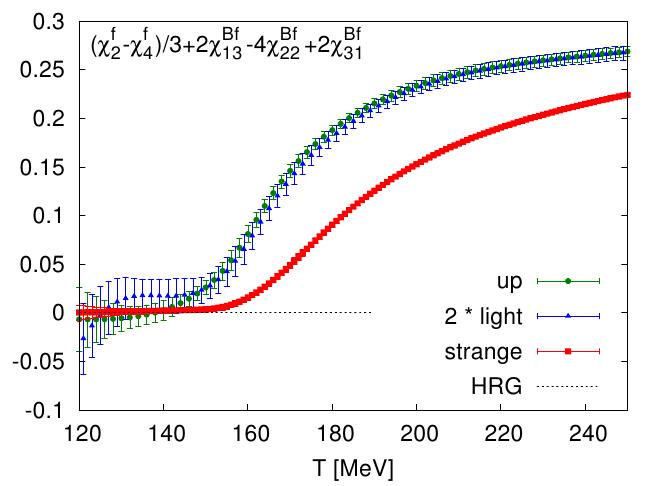}
\caption{The flavor hierarchy of the QCD transition: strange degrees of freedom ``dehadronize'' at a 
higher temperature than the light \cite{Bellwied:2013cta}.}
\label{fi:hier}
\end{center}
\end{figure}

An interesting observation was reported by Borsanyi \cite{Borsanyi:2013aya}. The
characteristic temperature values of fluctuations, that are related to strange
quarks, are typically $\sim 15$ MeV higher, than those related to light quarks.
This is demonstrated on Figure \ref{fi:hier}, where beside the already
defined $v_2$ the authors also plot 
\begin{align*}
(\chi_2^u - \chi_4^u)/3 + 2\chi_{13}^{Bu} -4 \chi_{22}^{Bu} + 2 \chi_{31}^{Bu},
\end{align*}
which is obtained by replacing the $\mu_S$-derivations in $v_2$ with
$\mu_u$-derivations. This latter is the chemical potential, that is coupled
to the number of up quarks.
The separation of the light and strange quark curves and
respective characteristic temperatures is evident from the plot. It can be seen in many
other fluctuation combinations, too. The authors call this finding as the flavor hierarchy of the QCD transition \cite{Bellwied:2013cta}.

\subsection{Ab-initio determination of freezout parameters}

The two central question of heavy-ion collision experiments are whether the
system created in these collision is thermal equilibrated, and if yes, what are
the corresponding parameters, ie. the temperature and chemical potential
values.

To obtain these parameters statistical hadronization models are widely used
(see eg. \cite{Andronic:2005yp}), in which the numbers of the outcoming hadrons
are fitted with simple Bose or Fermi distributions.  The so obtained
temperature and chemical potential values are called freezout parameters, below
this temperature one expects no change in the number and type of particles. A
problem with this approach is, that the typical freezout temperature turns out
to be $T_f\sim160$ MeV, which is inside the transition region of QCD, where the
use of a non-interacting hadronic description is questionable. 

Last year it has been realized, that conserved charge fluctuations can solve
this problem and using lattice QCD one can determine freezout parameters in
an ab-initio way \cite{Bazavov:2012vg}. Let us first consider the experimental
side, where we start with two lead ions, each of which carries $B=207$, $Q=82$
and $S=0$. After the collision the outgoing particles will also have the same
total charges, since these are conserved quantum numbers (considering only the
strong interaction).  So what exactly fluctuates? In order to get fluctuations,
one has to consider a subsystem, ie. particles coming only from a small part of
the system.  This is defined by imposing kinematical constraints on the
outgoing particles. There is no constraint any more on the charges of a
subsystem, charges can flow in and out, the measured values will be different
from one event to the other. These are the event-by-event fluctuations
\cite{Stephanov:1999zu}.

By relating event-by-event fluctuations to the fluctuations measured on the
lattice one can answer the two central questions of heavy-ion collisions:

\begin{enumerate}

\item The four unknown parameters, T, $\mu_B$, $\mu_Q$ and
$\mu_S$, can be determined by using four conditions: we require, that for four
observables the experimental and lattice values be equal. The so obtained parameters are called
ab-initio freezout parameters. Ab-initio, since nothing but the QCD
Lagrangian is assumed.  Freezout, since they reflect the state
of the system, after which there was no change in the number of charges.

\item Measuring observables other than the previous four can be used to decide
the question about the equilibrium. If for other fluctuations the experimental
and lattice values are still equal, then the equilibrium hypothesis gets more
support. If there is a discrepancy, then the fluctuations cannot have a thermal
origin.

\end{enumerate}

Additionally one can also plot the freezout parameters for different
experiments onto the temperature-chemical potential plane and see how it is
related to the QCD transition line.  This can be used to design fluctuation
based signatures for experiments \cite{Stephanov:2008qz}, in order to ease the
location of the QCD endpoint on the transition line (if it exists).

\begin{figure}
\begin{center}
\includegraphics[width=0.60\textwidth]{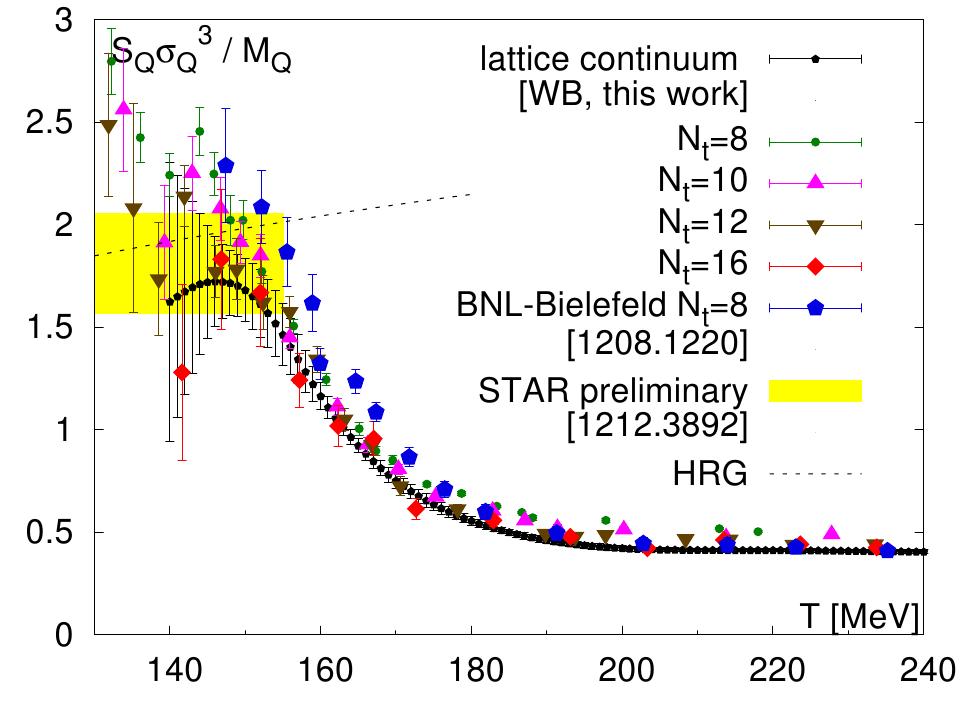}\\
\includegraphics[width=0.60\textwidth]{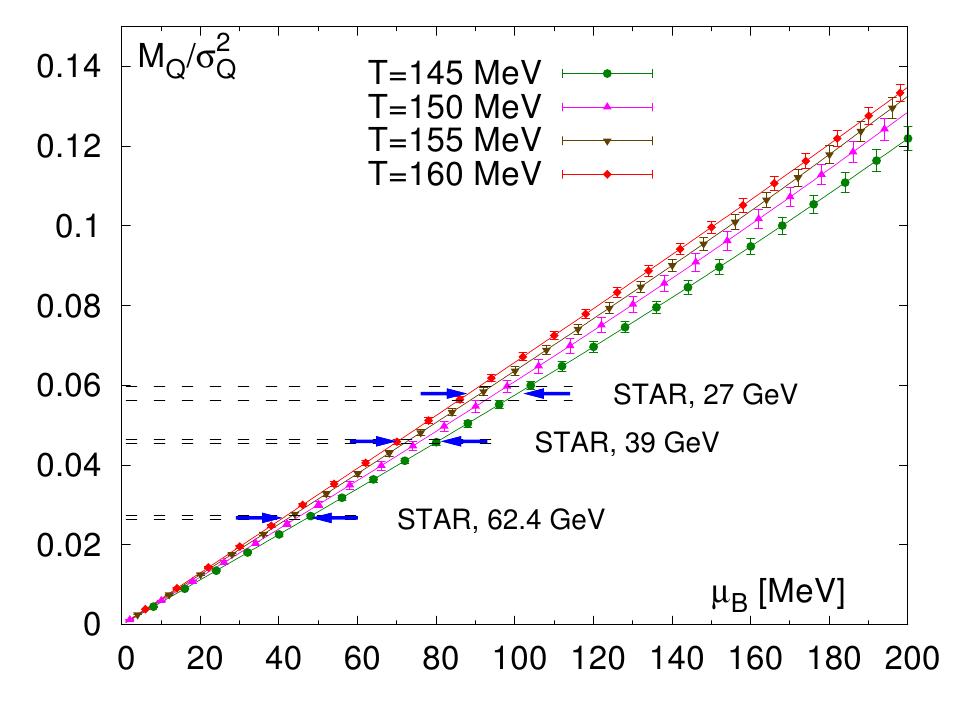}
\caption{By matching experimental and lattice data for the thermometer (left) and baryometer (right) one can extract
the freezout temperature and chemical potential \cite{Borsanyi:2013aya}.}
\label{fi:tmbm}
\end{center}
\end{figure}

The BNL-Bielefeld collaboration has proposed a concrete way to extract the
freezout parameters by matching experimental and lattice data \cite{Bazavov:2012vg}. Using
\begin{align*}
\langle S \rangle = 0 \quad \quad \text{and} \quad \quad \frac{\langle Q\rangle}{\langle B\rangle}= \frac{82}{207}
\end{align*}
one can express $\mu_S$ and $\mu_Q$ as the function of the other two parameters: $T$ and $\mu_B$. To determine
these two one chooses two fluctuations, the so-called thermometer and the baryometer, for which it is required, that their
experimental and lattice values be equal. A choice, that is convenient both from experimental and lattice point of view, is 
based on charge fluctuations:
\begin{align*}
\langle \delta Q^3 \rangle/\langle Q \rangle & \text{ as thermometer and}\\
\langle Q \rangle/\langle \delta Q^2 \rangle & \text{ as baryometer,}
\end{align*}
where $\delta Q= Q-\langle Q \rangle$.  Note, that in principle the
fluctuations also depend on the size of the system; this dependence is linear,
if the system is large enough. In order to cancel this unknown factor, it is
advantageous to work with fluctuation ratios.  The WB collaboration has
recently presented high-precision continuum extrapolated data for these two
observables \cite{Borsanyi:2013aya}, on Figure \ref{fi:tmbm} we use them to to illustrate the temperature and
chemical potential determination.  On the upper panel the thermometer is shown
as the function of temperature. The yellow band is the experimentally measured
value, the black points are the lattice results. From the equality of the two
we get an upper bound on the freezout temperature $T_f\lesssim 157$ MeV. The
lower panel shows the baryometer as the function of the chemical potential. The
horizontal lines are the experimental values for different beam energies,
equating with the lattice result yields eg. $\mu_f\gtrsim 95$ MeV for the smallest beam energy
on the plot.

\begin{figure}
\begin{center}
\includegraphics[width=1.0\textwidth]{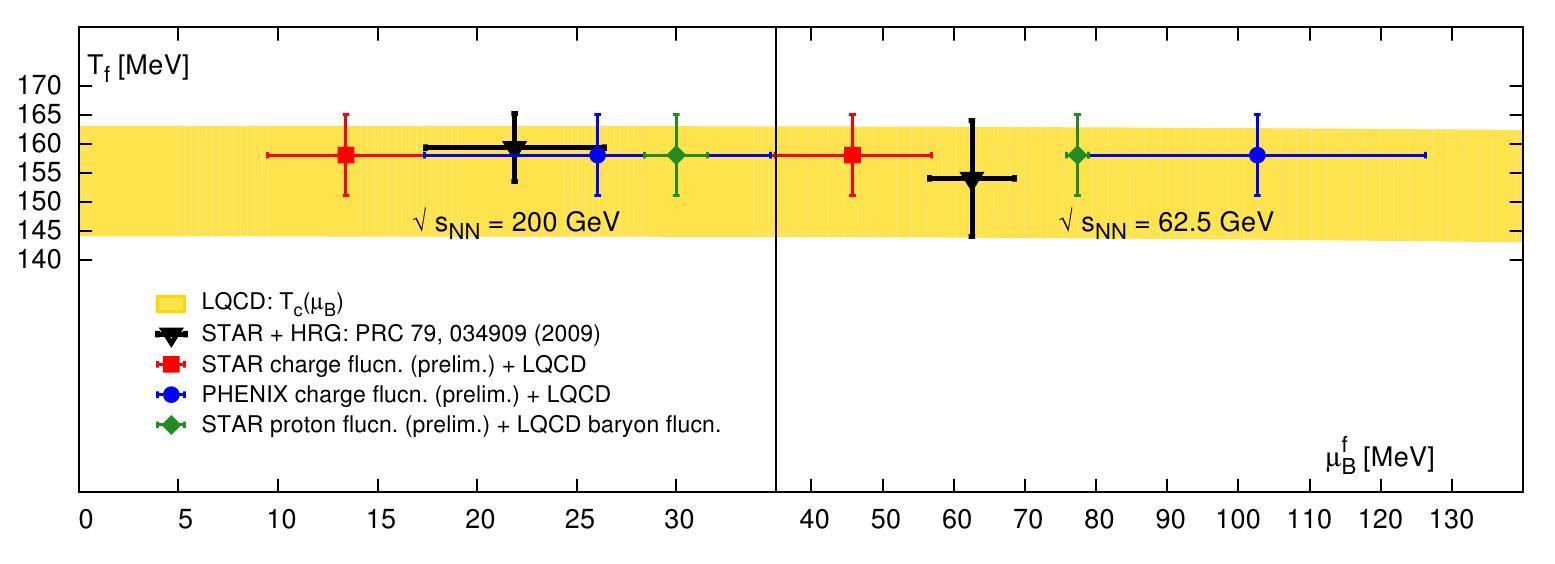}
\caption{Freezout parameters and the QCD transition line on the $\mu_B-T$ plane (plot from Wagner's talk).}
\label{fi:bnl}
\end{center}
\end{figure}
Wagner presented a plot of the BNL-Bielefeld collaboration featuring the
extracted freezout parameters and the QCD transition line on the $\mu_B-T$
plane. This is shown on Figure \ref{fi:bnl}. There are two different beam energies
corresponding to the left and right panels of the plot.  For each beam energy
there are four different freezout parameter determinations: the black is the
old approach based on statistical hadronization models. The colored ones are
the newly proposed ab-initio freezout parameter determinations: the red and
blue ones are based on charge fluctuations, the greens are based on matching
lattice baryon number fluctuations to the experimental proton number
fluctuations (note, that a matching using baryon number would require the
detection of neutral baryons, which is an uneasy task). There is some
disagreement between the different determinations, which is partly due to the
preliminary status of the experimental data, partly to the mismatch of baryon
and proton numbers. All freezout parameters are nicely consistent with the QCD
transition line (yellow band), suggesting that the freezout takes place at the
transition.

The main message of this section, that fluctuations provide a model-independent
way to extract freezout parameters in experiments and the first attempts have
already been taken.

\newpage 
\section{Columbia-plot}

\begin{figure}
\begin{center}
\includegraphics[width=0.60\textwidth]{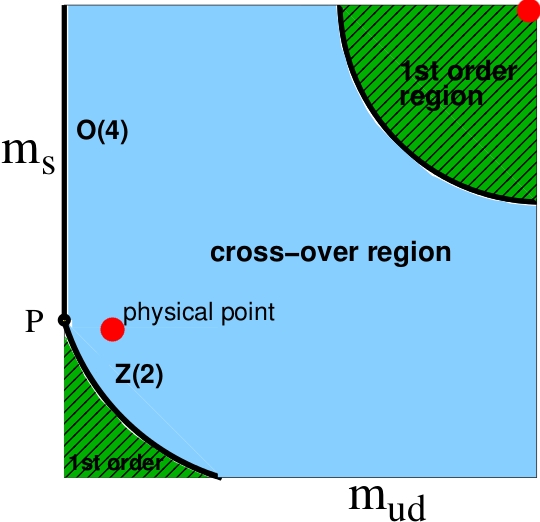}
\caption{The Columbia-plot: the order of the transition as the function of the light and strange
quark masses. Only the two red points are known for sure.}
\label{fi:col}
\end{center}
\end{figure}
A traditional topic in lattice thermodynamics is the order of the transition in
the chiral limit with various number of flavors. The status is usually
summarized in form of the Columbia-plot \cite{Brown:1990ev}, Figure \ref{fi:col}, which shows the order as the
function of the light and strange quark masses. There are only two points,
which are known with high confidence: the physical point is a crossover and the
pure SU(3) theory has a first order transition. The rest of the plot is a
prediction of an effective field theory, first advocated by Pisarski and
Wilczek \cite{Wilczek1,Wilczek:1992sf}. 

\subsection{Where is the first order transition in $n_f=3$?}

In case of three degenerate flavors the effective theory has a first-order
transition. If this also holds in QCD, then there should be a non-zero pion
mass at which the first order transition turns into crossover. The search of
this critical pion mass $m_\pi^c$ has already a long history, many simulations
have been carried out with staggered fermions.  Decreasing the lattice spacing
and/or improving the lattice action decreases the critical pion mass. The
current best searches were not even able to find a first order region, only an upper
bound on $m_\pi^c$ was possible:
\begin{center}
\begin{tabular}{|c|c|c|c|}
\hline
$N_t$ & action & $m_\pi^c$[MeV] & reference \\
\hline
4 & unimproved  & 260 & \cite{Karsch:2001nf,deForcrand:2007rq} \\ %mpi/Tc= 1.68 + Tc=154 MeV for conversion (Karsch's 3flavor)
6 & unimproved  & 150 & \cite{deForcrand:2007rq} \\ %mpi/Tc= 0.95
4 & p4          &  70 & \cite{Karsch:2003va} \\ %mpi= 68
6 & stout       & $\lesssim 50$ & \cite{Endrodi:2007gc} \\ %mpi= sqrt(0.12)*135 MeV
6 & HISQ        & $\lesssim 45$ & \cite{Ding:2011du} \\ %mpi= sqrt(0.10)*135MeV
\hline
\end{tabular}
\end{center}
A serious shortcoming of all these staggered studies is, that although the
pseudo Goldstone pion mass could be decreased at a fixed $N_t$, the root
mean squared pion mass practically does not change below some quark mass. In
the above studies it was always $m_\pi^{RMS}\gtrsim 400$ MeV.

A study with $n_f=3$ Wilson fermions was presented by Nakamura, the critical
pion mass was found to be at $m_\pi^c\sim500$ MeV. If it holds, it would mean,
that the negative result of the staggered simulations really lies in the too
large RMS pion mass: a slap in the face of these simulations.

\subsection{The role of the axial symmetry in the $n_f=2$ case}

In case of two degenerate flavors even the effective theory prediction becomes
ambiguous, it depends on the strength of the $U(1)_A$ axial anomaly at the
temperature of the $SU(2)_L\times SU(2)_R$ chiral restoration. If the anomaly is
strong enough, then the transition is second order in the $O(4)$ universality
class. If the anomaly is weaker, then one might get a first-order transition.
In case of a complete restoration even another second order transition, now
in the $U(2)_L\times U(2)_R/U(2)_V$ universality class, is possible \cite{Basile:2005hw}.

One way to study the restoration of the axial symmetry is to
measure the difference of the pseudoscalar ($\pi \sim
\overline{u}\gamma_5d$) and scalar ($\delta \sim \overline{u} d$) correlators:
\begin{align*}
\langle \pi(x)\pi(0) \rangle - \langle \delta(x) \delta(0) \rangle.
\end{align*}
Another approach is to look at the near-zero eigenmodes of the Dirac-operator,
their absence should also be a signature of the axial symmetry restoration.
There are several groups investigating along these lines partly with
contradicting results:
\begin{center}
\begin{tabular}{|c|c|c|l|}
\hline
$U(1)_A$ restored? & fermion & lattices & group,presenter \\
\hline
no  & domain-wall          & $8\cdot\{16,24,32\}^3$  & hotQCD \cite{Buchoff:2013nra}, Schroeder \\
no  & overlap on staggered & $8\cdot32^3$  & Bielefeld, Sharma \cite{Sharma:2013nva} \\
yes & overlap              & $8\cdot16^3$  & JLQCD \cite{Cossu:2013uua}, Taniguchi \cite{Aoki:2013zfa}\\
yes & domain-wall          & $6\cdot16^3$  & TW-QCD, Chiu \cite{Chiu:2013wwa}\\
\hline
\end{tabular}
\end{center}
For a definite conclusion one would like to have extrapolations to infinite
volume, chiral limit and continuum limit, none of the calculations can provide
all three.

\subsection{Direct determinations suggest second order for $n_f=2$}

\begin{figure}
\begin{center}
\includegraphics[width=0.55\textwidth]{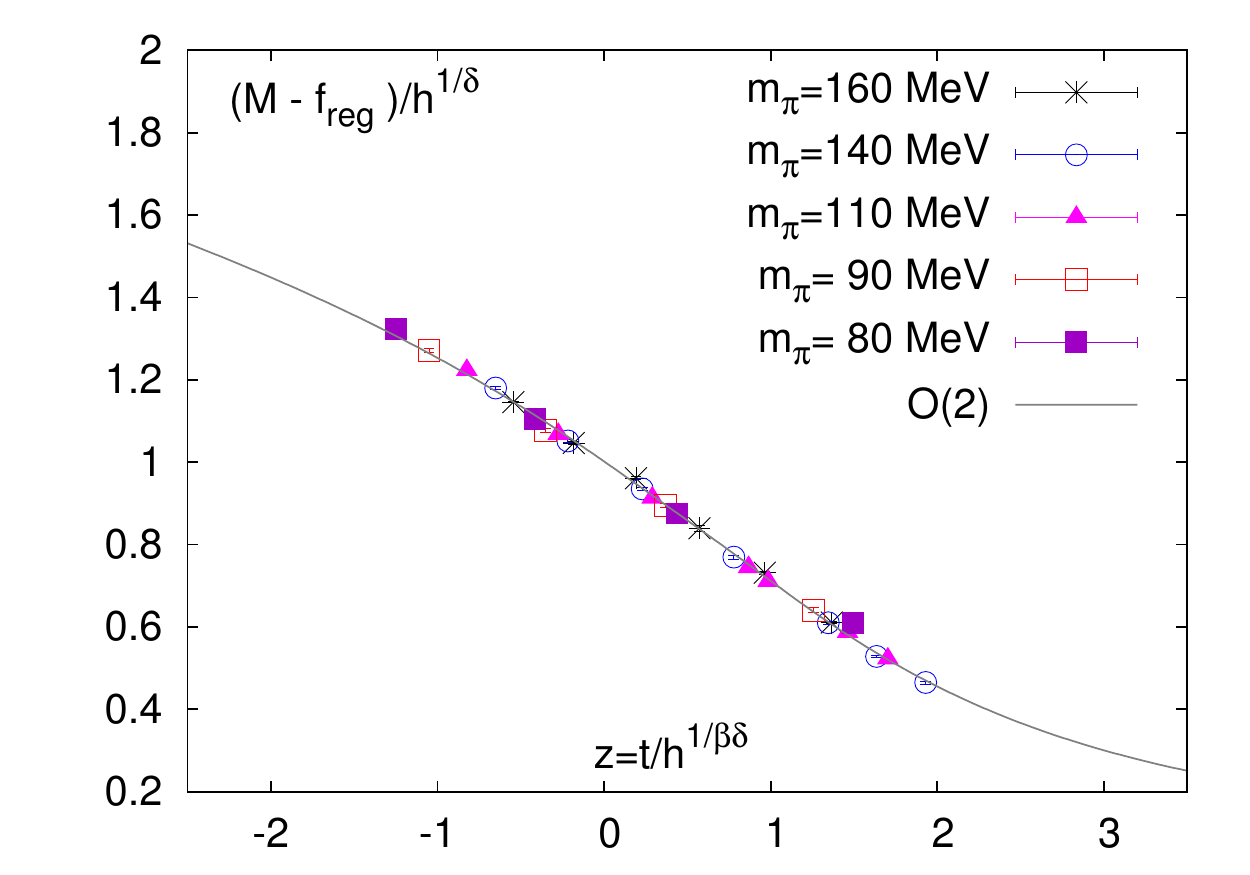}
\caption{
Scaling plot of the chiral condensate assuming,
that the the transition in second order in the two flavor chiral limit \cite{Ding:2013lfa}.}
\label{fi:o4}
\end{center}
\end{figure}

The current best evidence for a second order transition comes from the
BNL-Bielefeld collaboration. Ding presented \cite{Ding:2013lfa} an update of the results with highly improved
staggered quarks at one lattice spacing down to a pseudo Goldstone pion mass of $m_\pi \sim 80$
MeV (it is actually a $n_f=2+1$ simulation with a physical strange quark). No
signal of a first order transition has been found. The chiral condensate for
different quark masses and temperatures nicely collapse to a single scaling
curve assuming, that the transition is $O(4)$ second order in the chiral limit
(see Figure \ref{fi:o4}). As it has already been mentioned, a shortcoming of this approach
is, that with staggered quarks it is questionable to carry out a chiral limit
before the continuum extrapolation.

\begin{figure}
\begin{center}
\includegraphics[width=0.55\textwidth]{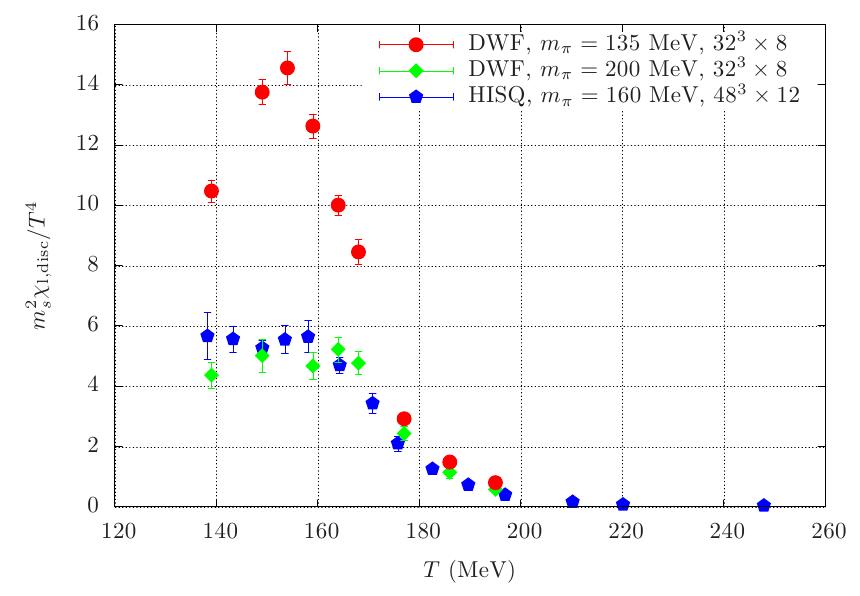}
\caption{QCD transition with domain wall fermions at the physical point, the plot shows the chiral susceptibility as the function
of the temperature \cite{Buchoff:2013nra}.}
\label{fi:dwf}
\end{center}
\end{figure}

There is an emerging segment of studies with non-staggered fermions. The most
impressive example to date is the work of the hotQCD collaboration \cite{Buchoff:2013nra}. Domain-wall
fermions are used on upto $8\cdot32^3$ lattices. The pion mass is gradually
decreased, there are runs even with physical pion mass; the strange mass is set
to the physical value. It has been decided to boost this even further: Schroeder presented
runs on a $8\cdot64^3$ lattice at $m_\pi \sim 100$ MeV!  No sign of a first
order transition has been found, yet.  As Figure \ref{fi:dwf} shows the peak of the chiral
susceptibility at the physical point is $T_c\sim 155$ MeV. This is the first
fully independent confirmation of the previously disputed staggered result (see
Section \ref{se:tc}).

The WB collaboration also investigates the transition with chiral fermions
(overlap fermion, \cite{Borsanyi:2012xf}). The main focus is on the continuum
limit, there are now four lattice spacings at a pion mass of $m_\pi\sim 320$
MeV. These results also support the universality: there is a nice agreement
with previous staggered data. There are also simulations with non-chiral
fermions: the tmfT collaboration uses twisted mass fermions
\cite{Burger:2013hia}, the Frankfurt-Mainz group Wilson fermions
\cite{Brandt:2013mba}. None of them are conclusive on the order of the
transition in the chiral limit, yet.

\subsection{Imaginary $\mu$ approach says first order in $n_f=2$ on a coarse lattice}

All previous approaches use the same strategy: simulate with gradually smaller
pion masses in the crossover region and hope, that at some pion mass the
transition turns into first order.  However there is not even a slightest clue
about the value of the critical pion mass.  The problem with this, that one
cannot write an honest proposal to get CPU time for such a project, because for
that one would have to know how small pion masses are to be simulated.

An interesting alternative approach, which might circumvent this problem,
arises by considering the phase diagram at imaginary chemical potential
$\mu_I$. The idea was discussed first by D'Elia and Sanfilippo \cite{D'Elia:2009qz}, it was
systematically developed by de Forcrand and Philipsen \cite{deForcrand:2010he}.  Instead of going down
with the pion mass at $\mu_I=0$ in the crossover region, one turns on the chemical
potential at a fixed pion mass. It can happen, that the transition gets
stronger and eventually turns into first order at a critical pion mass
$m_\pi^c(\mu_I)$. If this point is found, then one just has to follow, what
happens with $m_\pi^c(\mu_I)$ as the chemical potential goes back to zero. This
strategy was successfully used to determine the $n_f=2$ order of the transition
on $N_t=4$ unimproved staggered lattices: it is first order \cite{Bonati:2013tqa}.
\begin{figure}
\begin{center}
\includegraphics[width=0.7\textwidth]{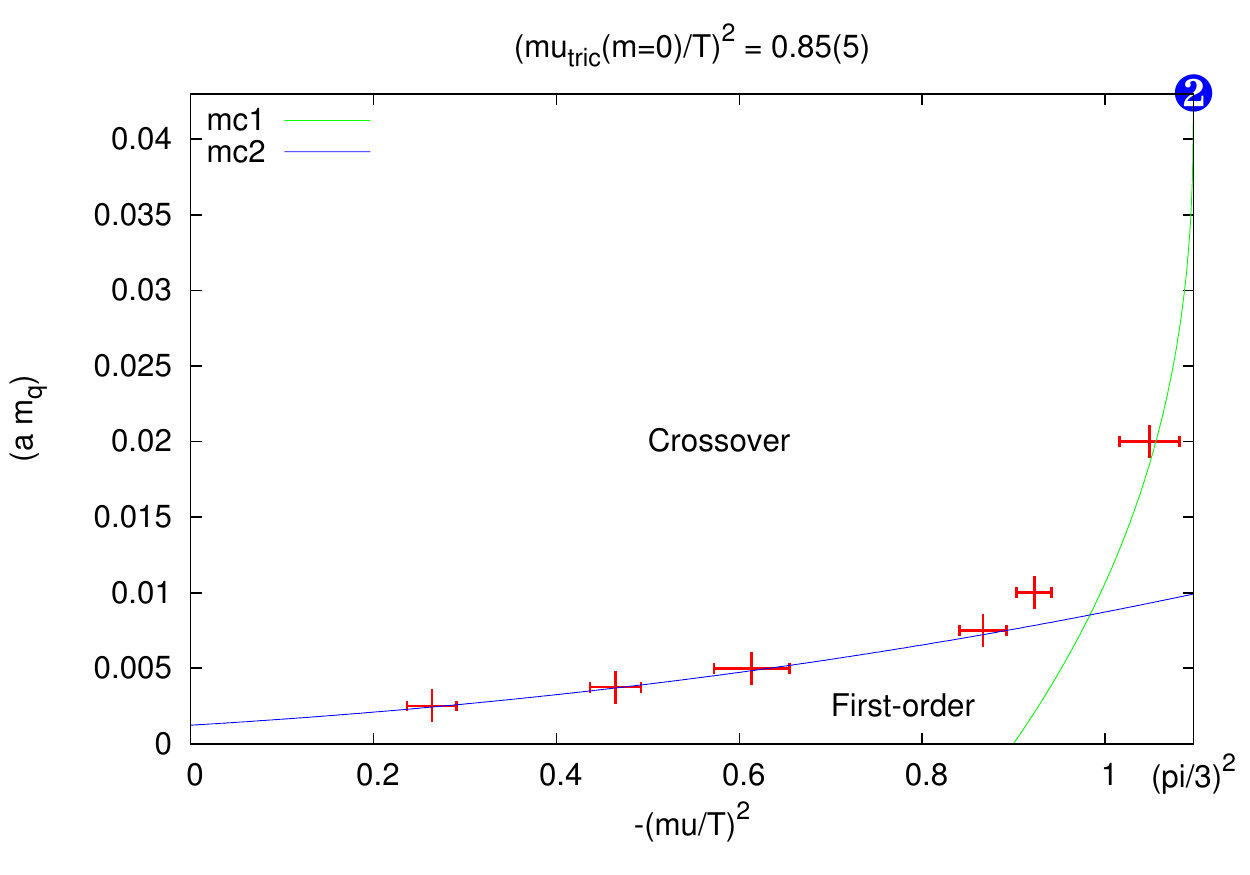}
\caption{Phase diagram in the imaginary $\mu$-quark mass plane for $n_f=2$. Imaginary $\mu$ simulations helped
to determine the order of transition in the chiral limit: it is first order \cite{Bonati:2013tqa}. Note, that the lattice is very coarse, $N_t=4$.}
\end{center}
\end{figure}

This result contradicts previous direct determinations with improved actions
(see previous subsection), which find no existence of a first order region.
What might seem a contradiction first, fits actually nicely into the effective
theory picture. Staggered fermions on coarse lattices have serious cutoff
effects in the axial symmetry breaking and a weak or non-existent axial term in
the effective theory actually favors a first-order transition. So the order of
the transition can easily be first order on coarse lattices without
improvement, and turn out to be second order on finer lattices and/or with
improvement. 

Ejiri presented an approach \cite{Ejiri:2013kfa}, which is similar in spirit.
Increasing the number of flavors also makes the transition stronger and the
same procedure can be applied as for imaginary $\mu$.

\section{Magnetic field}

The rising star among the topics of recent years is lattice QCD at finite
magnetic field $B$. It has gained momentum after realizing, that spectator
particles in non-central heavy-ion collisions produce extreme large magnetic
fields ($\sim 10^{15} T$). This field induces a charge current, if there is an
imbalance between left and right handed particles, this is the chiral magnetic
effect, \cite{Kharzeev:2007jp}.  This can be used to test the non-Abelian
nature of the strong interaction.  On the lattice on a fixed instanton
background the chiral magnetic effect can be nicely demonstrated
\cite{Abramczyk:2009gb}.  On ``real'' configurations the effect is more
complex, it has been studied by Buividovich et al \cite{Buividovich:2009wi},
Yamamoto \cite{Yamamoto:2011gk} and in two color QCD by Ilgenfritz et al \cite{Ilgenfritz:2012fw}.

\begin{figure}
\begin{center}
\includegraphics[width=0.54\textwidth]{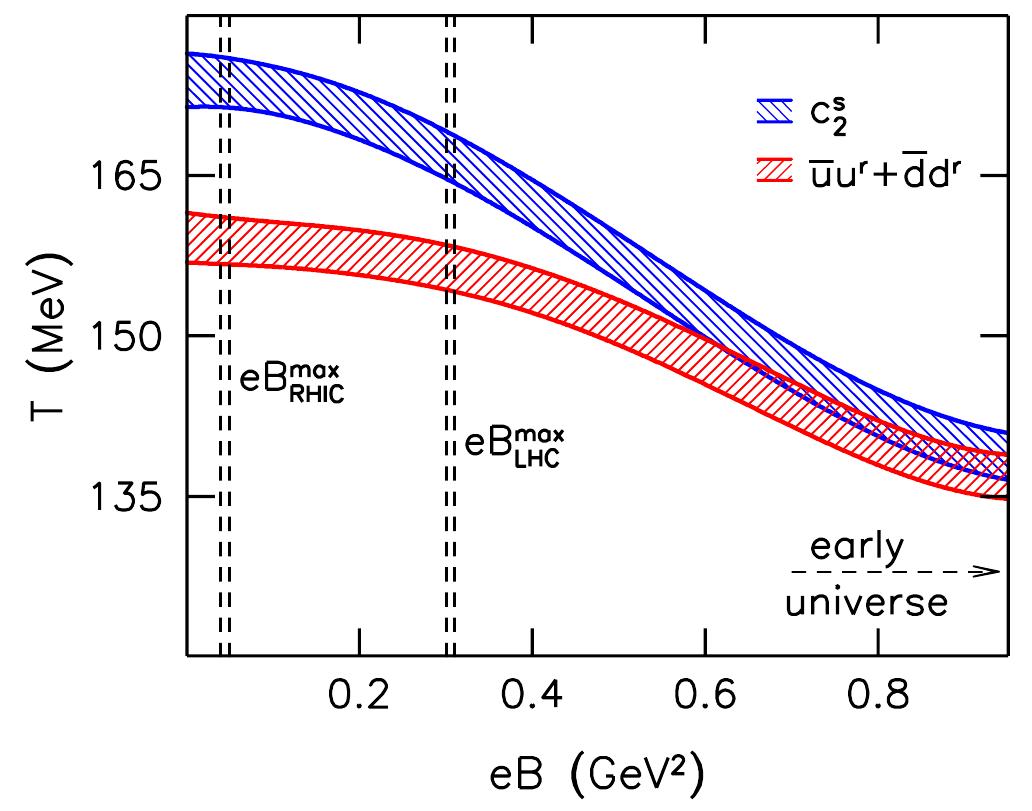}
\caption{Transition temperatures, defined from two different observables, as the function of the magnetic field.}
\label{fi:mag}
\end{center}
\end{figure}
Thermodynamical properties at finite $B$ have been extensively
studied. The common wisdom about the transition temperature was, that $T_c$
increases with $B$. This was even supported by first lattice studies by
D'Elia et al \cite{D'Elia:2010nq}. The continuum extrapolation, carried out by Bali et al \cite{Bali:2011qj}, has
changed the picture completely, the result is shown on Figure \ref{fi:mag}: $T_c$ gets
smaller with increasing $B$. The order of the transition has also been
investigated: there is no volume dependence in the chiral susceptibility, also
the width of the peak stays constant with $B$. The transition remains therefore a
crossover upto the largest magnetic field that was studied so far on the lattice
$\sqrt{eB}\sim 1$ GeV.

\begin{figure}
\begin{center}
\includegraphics[width=0.58\textwidth]{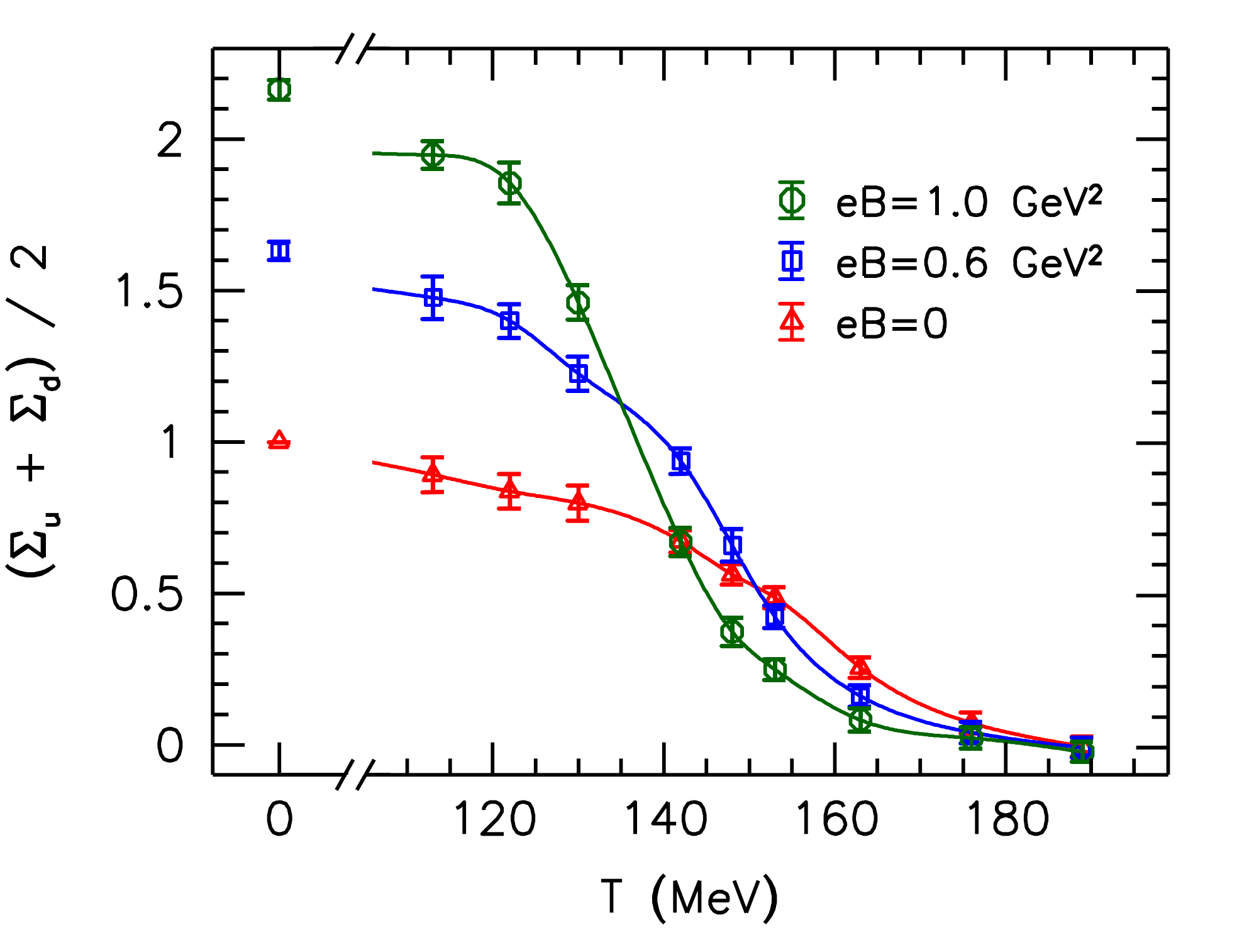}
\caption{Temperature dependence
of the chiral condensate showing the inverse magnetic catalysis around the transition \cite{Bali:2012zg}.}
\label{fi:mag2}
\end{center}
\end{figure}

The common wisdom about the transition temperature can actually traced back to
the behaviour of the chiral condensate $\overline{\psi}\psi$ at finite $B$.
Previously it was expected, that a magnetic catalysis takes place for all
temperatures, which means, that the $\overline{\psi}\psi$ increases with $B$.
As current lattice studies show (see Figure \ref{fi:mag2}), this is indeed the case for small
temperatures, however around the transition an opposite effect, the inverse
magnetic catalysis can be observed: $\overline{\psi}\psi$ decreases with $B$ \cite{Bali:2012zg}.
This then results in the decrease of the transition temperature. A closer look
at magnetic catalysis and inverse magnetic catalysis was presented by
Kovacs \cite{Bruckmann:2013oba}. The magnetic field dependence of the chiral condensate
comes from two sources \cite{D'Elia:2011zu}:
\begin{align*}
\langle \overline{\psi}\psi \rangle \sim \int \text{Tr} D^{-1}(B) \cdot \det D(B).
\end{align*}
There is a ``valence'' contribution $\text{Tr} D^{-1}(B)$, which indeed always
increases with $B$ due to the increase in the eigenvalue density of the
Dirac-operator on a fixed gauge background. This was known even before lattice
studies. However there is also a ``sea'' contribution: it takes into account the change
in $\overline{\psi}\psi$ due to the change of the typical gauge backgrounds. It
can both increase or decrease with the magnetic field.  This second
contribution was not taken into account in previous qualitative analyses, it
has been first caught in lattice QCD simulations. It can actually compensate
the catalytic effect of the valence contribution yielding an inverse magnetic
catalysis in total \cite{Bruckmann:2013oba}.

An interesting question is, what would a piece of quark-gluon plasma do, if we
approached it with a magnet? Would it be attracted or repelled by the magnet? Is
it a para- or a diamagnetic material? The answer can be given by calculating
the sign of the magnetic susceptibility:
\begin{align*}
\xi(T)= \frac{T}{V} \left.\frac{\partial^2}{\partial (eB)^2}\right|_{0} \log Z
=\begin{cases}
>0 & \text{para} \\
<0 & \text{dia} \\
\end{cases}
\end{align*}
The renormalization of this quantity was given in Appendix A of Reference
\cite{Bali:2013esa} and it follows that at zero temperature $\xi(0)=0$.  The
derivative is somewhat cumbersome on the lattice, since $B$ can only be changed
in discrete steps on a periodic lattice \cite{AlHashimi:2008hr}.  There are
three groups calculating $\xi(T)$ at finite temperature, in three different
ways.  Bonati presented a technique using a finite difference method
\cite{Bonati:2013lca}. DeTar was showing results using a non-conventional definition
of the lattice $B$ field combined with Taylor-expansion \cite{Levkova:2013qda}. Endrodi utilized two
approaches to obtain the $B$ dependence of $\log Z$ \cite{Bali:2013txa}: one based on calculating
anisotropies, the other on a novel type of integral method.  All
yield the same result: the quark-gluon plasma is a paramagnetic medium. This paramagnetic
property might increase the elongation of the plasma produced in non-central heavy ion collisions \cite{Bali:2013owa}.

\section*{Acknowledgement}

The author would like to thank Gergo Endrodi for help with preparing the
magnetic field summary, the Debrecen and Pisa groups for hospitality and many
colleagues for sending results: Alexei Bazavov, Claudio Bonati, Bastian Brandt,
Ting-Wai Chiu, Massimo d'Elia, Carleton DeTar, Heng-Tong Ding, Tamas Kovacs,
Ludmila Levkova, Michael Mueller-Preussker, Yoshifumi Nakamura, Francesco
Negro, Chris Schroeder, Sayantan Sharma, Mathias Wagner.

\end{document}